\documentclass[fleqn,twoside]{article}
\usepackage{espcrc2}
\usepackage{amssymb}
\usepackage[figuresright]{rotating}
\newenvironment{Eqnarray}%
          {\arraycolsep 0.14em\begin{eqnarray}}{\end{eqnarray}}

\newcommand{\bc}{\begin{center}}
\newcommand{\ec}{\end{center}}
\newcommand{\eq}{\begin{equation}}
\newcommand{\ee}{\end{equation}}
\newcommand{\ea}{\begin{Eqnarray}}
\newcommand{\eea}{\end{Eqnarray}}
\newcommand{\be}{\begin{equation}}
\newcommand{\bea}{\begin{eqnarray}}

\newcommand{\AmS}{{\protect\the\textfont2A\kern-.1667em\lower.5ex\hbox{M}\kern-.125emS}}

\title{
\thispagestyle{empty}
\vspace{-25mm}
\rightline{\small DESY 04-166~~~~~}
\rightline{\small ITEP-LAT/2004-17~~~~~}
\rightline{\small KANAZAWA-04-12~~~~~}
\vspace{10mm}
Finite temperature QCD with two flavors of dynamical quarks on $24^3 \times 10$ lattice
\thanks{Talk given by Y. N. at Lattice'04.}
}
\author{Y. Nakamura\address{Kanazawa University, Kanazawa 920-1192, Japan\\[-0.5em]}, 
V.G. Bornyakov$\address{Institute for High Energy Physics, RU-142284 Protvino, Russia\\[-0.5em]}, 
{\rm M.N. Chernodub}$\address{ITEP, B.Cheremushkinskaya 25, RU-117259 Moscow, Russia\\[-0.5em]}, 
Y. Mori$^{\rm a}$, 
S.M. Morozov$^{\rm c}$, 
M.I. Polikarpov$^{\rm c}$, \\
G. Schierholz\address{NIC/DESY Zeuthen, Platanenallee 6, D-15738 Zeuthen, Germany\\[-0.5em]}, 
A.A. Slavnov\address{Steklov Mathematical Institute, Vavilova 42, RU-117333 Moscow, Russia\\[-0.5em]}, 
H. St\"uben\address{Konrad-Zuse-Zentrum f\"ur Informationstechnik Berlin, D-14195 Berlin, Germany\\[-0.5em]} 
and T. Suzuki$^{\rm a}$
}

\begin{document}
\begin{abstract}
We present results obtained in QCD with two flavors of non-perturbatively improved Wilson
fermions at finite temperature on $16^3 \times 8$ and $24^3 \times 10$ lattices.
We determine the transition temperature in the range of quark masses $0.6<m_\pi/m_\rho<0.8$ at
 lattice spacing a$\approx$0.1 fm and extrapolate the transition temperature to the continuum
and to the chiral limits.
\end{abstract}
\maketitle

\section{INTRODUCTION}
\vspace{-1mm}
In order to obtain predictions for the real world from lattice QCD, we have to
extrapolate the lattice data to the continuum and to the chiral limits.
Recently the Bielefeld group~\cite{kpe} and the CP-PACS
collaboration~\cite{aakcp} using different fermion actions obtained consistent
values for the critical temperature $T_c$ in the chiral limit, 
albeit on rather coarse lattices at $N_t=4$
and 6. Edwards and Heller~\cite{eh} determined $T_c$ for $N_t=4$, 6 using
nonperturbatively improved Wilson fermions. We compute $T_c$ on finer
lattices with $N_t=8$ and 10 with high statistics. Our results for $N_t=8$ 
were reported in Ref.~\cite{previous}.

\section{SIMULATION}
\vspace{-1mm}
We use fermionic action for non-perturbatively improved Wilson fermions:
\ea
S_F &=&  S^o_F - \frac{\rm i}{2} \kappa\, g\,
c_{sw} a^5
\sum\nolimits_x \bar{\psi}(x)\sigma_{\mu\nu}F_{\mu\nu}\psi(x) \nonumber,
\eea
where $S^o_F$ is the original Wilson action, $c_{sw}$ was calculated in
\cite{Jansen:1998mx}.

\begin{figure}[!hbt]
\centerline{\includegraphics[angle=0,scale=0.25,clip=true]{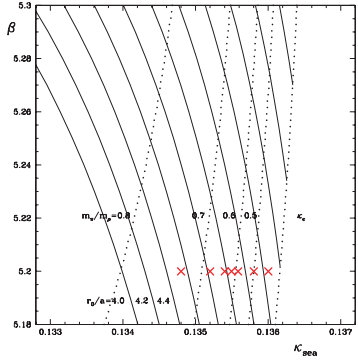}}
\vskip -11mm \caption{The lines of constant $r_0 /a$ and $m_\pi /m_\rho$ at
$T=0$. Crosses correspond to parameters used in simulations of $24^3 \times 10$ lattice.} 
\label{constant_DIK.ps}
\vskip -6mm
\end{figure}

Configurations are generated on $16^3 \times 8$ ($\beta=5.2$ and $5.25$) and
$24^3 \times 10$ ($\beta=5.2$) lattices at various $\kappa$.
The values of $\kappa$ and the corresponding number of configurations for
$16^3 \times 8$ and $24^3 \times 10$ lattices can be found in Ref.~\cite{previous} 
and Table 1, respectively.
The number of configurations for $24^3 \times 10$ lattice is not large
enough and results for this lattice are preliminary.
We use results obtained at T=0 to fix the scale.
The contour plot of lines of constant $r_0/a$ and $m_\pi/m_\rho$~\cite{lambda} 
is shown in Fig.~\ref{constant_DIK.ps}.
\vskip -1.5mm
{\footnotesize
\bc
\begin{tabular}{|c|c|c|c|c|} \hline
$\kappa$ & $0.1348$ & $0.1352$ & $0.1354$ & $0.1355$ \\ \hline
\# conf. & $   678$ & $   679$ & $ 1,234$ & $   799$ \\ \hline \hline
$\kappa$ & $0.1356$ & $0.1358$ & $0.1360$ & $      $ \\ \hline
\# conf. & $ 2,429$ & $   480$ & $   617$ & $      $ \\ \hline
\end{tabular}\\
\vskip 0.5mm
{\normalsize{Table 1:{Simulation statistics on $24^3 \times 10$.}}}
\ec
}

\section{CRITICAL TEMPERATURE}
\vspace{-1.5mm}
We use the Polyakov loop susceptibility to determine the transition
point. The critical value of $\kappa$ turns out to be $\kappa_t$=0.1354. Using
$r_0$=0.5 fm and interpolating $r_0/a$ to the critical point, we obtain for
the crtical temperature:
\bc
$T_c = 196(4)$MeV,~~~${m_\pi /m_\rho} = 0.64(3)$
\ec
\begin{figure}[!thb]
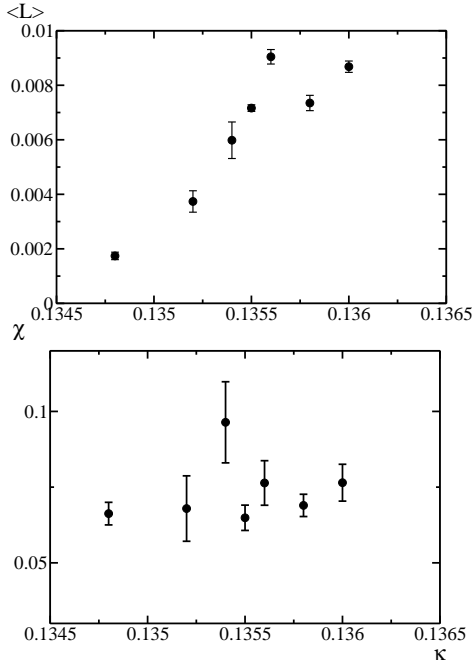

\begin{center}
\includegraphics[angle=0,scale=0.24,clip=true]{polyakov.loop.nonabelian.2410.eps}
\vskip -3mm
\includegraphics[angle=0,scale=0.24,clip=true]{susceptibility.nonabelian.2410.eps}
\end{center}
\vskip -13mm
\caption{Polyakov loop (top) and its susceptibility (bottom) at $\beta=5.2$ on $24^3 \times 10$.}
\label{NA}
\vskip -3mm
\end{figure}

\section{CONTINUUM AND CHIRAL LIMITS}
At small enough lattice spacing and quark mass one can extrapolate the critical
 temperature $T_c$ to the continuum and the chiral limits using formula:
\[
 T_c r_0 = T_c^{m_q , a \to 0} r_0 +C_a ({a \over r_0})^2
 +C_q ({1 \over \kappa} - {1 \over \kappa_c})^{{1 \over \beta\delta}} \,,
\]
where $T_c^{m_q , a \to 0}$ corresponds to the
extrapolated value of the critical temperature and $\beta$ and $\delta$ are critical indices.

We make an attempt to fit four values for $T_c r_0$ (see Table~3), obtained
at rather large quark masses, to estimate the parameters in this extrapolation
expression.

{\footnotesize
\bc
\begin{tabular}{|c|c|c|c|c|l|} \hline
$T_c r_0 $&$a/r_0   $&$\beta$&$\kappa_t$&$L_t$&                      \\  \hline
$0.50(1) $&$0.201(4)$&  5.2  & 0.1354  & 10  & prelim.              \\  \hline
$0.53(1) $&$0.234(4)$&  5.2  & 0.1345  &  8  & Ref.~\cite{previous} \\  \hline
$0.56(1) $&$0.225(4)$&  5.25 & 0.1341  &  8  & Ref.~\cite{previous} \\  \hline
$0.57(2) $&$0.29(1) $&  5.2  & 0.1330  &  6  & Ref.~\cite{eh}       \\  \hline
\end{tabular}\\
\vskip 2mm
{\normalsize{Table 3:Available data for $T_c r_0$.}}
\ec
\bc
\begin{tabular}{|c|c|c|c|c|c|} \hline
${1 \over \beta\delta}$&$T_c r_0  $&$C_a     $&$C_q    $&$\chi^2/dof  $&order \\ \hline
$ 0.54                $&$ 0.44(2) $&$-0.9(5) $&$ 0.5(1)$&$  0.26      $&  2nd \\ \hline
$ 1                   $&$ 0.51(3) $&$-1.3(7) $&$ 0.9(2)$&$  0.13      $&  1st \\ \hline
\end{tabular}
\vskip 2mm
{\normalsize{Table 4:Fitting results.}}
\ec}

\begin{figure}[!htb]
\centerline{\includegraphics[angle=0,scale=0.36,clip=true]{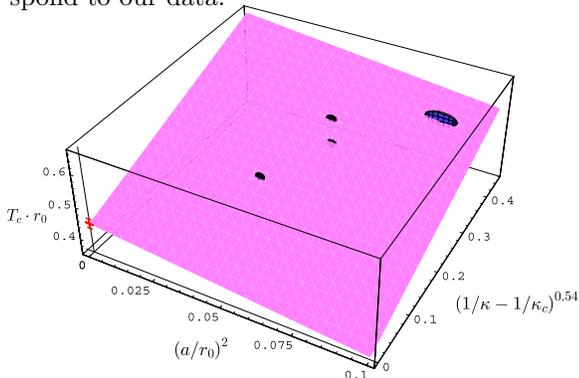}}
\vskip -11mm
\caption{ $T_c r_0$ as a function of lattice spacing and quark mass
if the second order phase transition is realized. Best fit is shown by the plane.}
\label{fit054}
\vskip -8mm
\end{figure}

\begin{figure}[!htb]
\centerline{\includegraphics[angle=0,scale=0.37,clip=true]{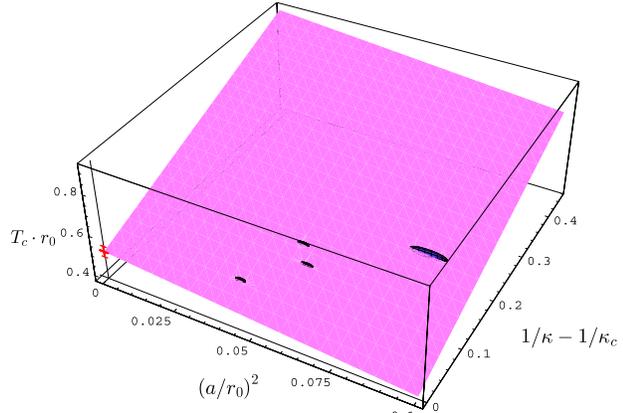}}
\vskip -11mm
\caption{ The same as in Fig.4 but for the first order phase transition.}
\label{fit100}
\vskip -8mm
\end{figure}

We extrapolate the value of the critical temperature
using different values of 0.54 and 1 as $1/\beta\delta$.
If the transition in two-flavor QCD is second order, the transition is
expected to belong to the universality class of the $3D$ $O(4)$ spin model
with  $1/\beta\delta$$\approx$0.54. If the transition is first order, then
$1/\beta\delta$=1. Table 4 and Figs.~\ref{fit054}, \ref{fit100} present
fitting results.
We get the critical temperature in the continuum and
in the chiral limits.

\vskip 2mm
\noindent
In the case of $1/\beta\delta$$=$0.54:
\eq\label{Tc054} T_c^{m_q , a \to 0} = 174(8)~\mbox{MeV}\,.\ee
This value agrees with values obtained in Refs.\cite{kpe,aakcp}.

\vskip 2mm
\noindent
In the case of $1/\beta\delta$=1:
\eq\label{Tc100} T_c^{m_q , a \to 0} = 201(12)~\mbox{MeV}\,.\ee

Although some lattice studies \cite{kpe,aakcp}  indicate second order
chiral transition in two-flavor QCD, there are also results \cite{excluO4}
supporting first order transition. Results of our fits do not allow to
discriminate between first and second order transitions  because of rather
large errors in $T_c r_0$ values.

\section{CONCLUSIONS}
\begin{figure}[!htb]
\centerline{\includegraphics[angle=0,scale=0.28,clip=true]{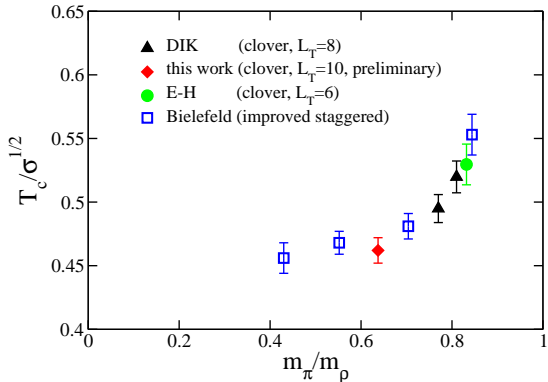}}
\vskip -11mm
\caption{$T_c/ \sqrt{\sigma}$ as a function of $m_{\pi} / m_{\rho}$.
The circle and squares show results of \cite{eh} and \cite{kpe} respectively.
The diamond and triangles correspond to our data.}
\label{tes}
\vskip -8mm
\end{figure}

We determined the critical temperature in full QCD on $24^3 \times 10$ lattice 
at $\beta =5.2$ with $N_f=2$ clover fermions using Polyakov loop susceptibility. 
Our results are in agreement with the results of other groups, as it is shown in
Fig.~\ref{tes}.
We extrapolate the critical temperature $T_c$ to the continuum and the chiral limits
both in cases of first and second order transition.
The extrapolation results are given by (\ref{Tc054}) and (\ref{Tc100}).
We are continuing simulations on $24^3 \times 10$ lattice
to get better precision of $T_c$ value on this lattice.

\vskip -8mm

\vskip -10mm

\section{ACKNOWLEDGEMENTS} 
This work is supported by the SR8000 Supercomputer Project of High Energy
Accelerator Research Organization (KEK). A part of numerical measurements has
been done using NEC SX-5 at Research Center for Nuclear Physics (RCNP) of Osaka
University. 
The numerical simulations of this work were done using RSCC computer clusters in 
RIKEN. We wish to acknowledge the support of the computer center at RIKEN. 
VGB, MNC and MIP are supported by grants RFBR 02-02-17308, 04-02-16079,
RFBR-DFG-03-02-04016, DFG-RFBR 436 RUS 113/739/0, INTAS-00-00111, CRDF
award RPI-2364-MO-02 and MK-4019.2004.2, A.A.S. is supported by grant
Scient.school grant 2052-2003.1.
T.S. is supported by JSPS Grant-in-Aid for Scientific Research on Priority 
Areas 13135210 and (B) 15340073.

\end{document}